\documentclass[draft]{aipproc}

\layoutstyle{6x9}

\begin{document}

\title{Burst Populations and Detector Sensitivity}

\author{David L. Band}
{address={Code 661, NASA/Goddard Space Flight Center,
Greenbelt, MD 20771}}

\begin{abstract}
The $F_T$ (peak bolometric photon flux) vs. $E_p$ (peak
energy) plane is a powerful tool to compare the burst
populations detected by different detectors.  Detector
sensitivity curves in this plane demonstrate which burst
populations the detectors will detect.  For example, future
CZT-based detectors will show the largest increase in
sensitivity for soft bursts, and will be particularly
well-suited to study X-ray rich bursts and X-ray Flashes.
Identical bursts at different redshifts describe a track in
the $F_T$-$E_p$ plane.
\end{abstract}

\maketitle

\section{Introduction}

Burst detectors are sensitive in different energy bands,
and some bursts emit most of their energy at 10~keV while
others emit at 1~MeV.  Here I present a methodology to
compare the burst populations that different detectors will
detect. The issue is what bursts will cause a detector to
trigger.

The standard rate trigger monitors a detector's count rate
for a statistically significant increase\cite{1}. The
counts are binned over an energy range $\Delta E$ and time
bin $\Delta t$, and the number of counts in this bin is
compared to the expected number of background counts.  For
example, BATSE used $\Delta E$=50--300~keV for most of the
{\it CGRO} mission and $\Delta t=$0.064, 0.256 and 1.024~s.
Usually the expected background is calculated from the
counts accumulated over a period (e.g., 20~s long) before
the bin being tested, and periodically (e.g., every 10~s)
the background is recalculated.  Most often the background
is modelled as a constant in time, but polynomials can be
fit.

A rate trigger tests the null hypothesis that only
background counts are present.  The test is whether the
probability that the observed number of counts in a $\Delta
E$-$\Delta t$ bin is a fluctuation is smaller than a
threshold value.  In the Gaussian limit the fluctuations
are measured in units of $\sigma$, the square root of the
expected number of counts, and the threshold fluctuation
probability can be mapped into the number of $\sigma$. Thus
the test is whether the observed number of counts exceeds
the expected number by more than a threshold multiple of
$\sigma$.

When more than one detector is present (e.g., BATSE's 8
detectors, the 12 NaI detectors of GLAST's GBM) the
requirement may be that more than one detector trigger.
Almost always a detector's sensitivity varies across the
field-of-view (FOV). If two detectors must trigger, the
spatial sensitivity depends on the angle to each detector.

For coded mask detectors such as HETE II's WXM and Swift's
BAT the rate trigger is followed by imaging, and a new
point source in the image must confirm the burst trigger.

\section{A Common Sensitivity Vocabulary}

A detector's sensitivity is the minimum count rate that
triggers the detector.  This corresponds to the burst's
peak count rate when integrated over $\Delta E$ and $\Delta
t$. Because bursts are not constant for seconds and burst
lightcurves differ at different energies, the sensitivity
will differ for different sets of $\Delta E$ and $\Delta
t$, and a smaller threshold count rate for one set of
$\Delta E$ and $\Delta t$ compared to another set may not
indicate that fainter bursts will be detected.  In
addition, because of the detector's finite energy
resolution, the count rate over $\Delta E$ is not the
photon flux integrated over $\Delta E$, and the count rates
for two detectors in the same nominal $\Delta E$ cannot be
compared directly.  This complicates the comparison between
different detectors and different triggers for the same
detector.  A common sensitivity measure is necessary; this
quantity will be burst-dependent.

The sensitivity measure should be a measure of the burst's
intensity that is independent of the detector or the
trigger specifics\cite{2}.  I propose $F_T$, the photon
flux integrated over 1--1000 keV and 1~s. This may not be
the most physically interesting or relevant intensity
measure, but it is closely related to the count rate that
triggers the detector.  Converting the count rate over
$\Delta E$ to $F_T$ requires the spectrum at the time of
the peak flux, but this spectrum is required for the
conversion from peak count rate to peak photon flux (e.g.,
BATSE's 50--300 keV peak flux). For a given detector
trigger (i.e., with a choice of $\Delta E$ and $\Delta t$)
$F_T$ will depend on the particulars of the burst: the
lightcurve around the time of the peak flux, and the
spectrum at this time.  Here I do not focus on the
dependence on the lightcurve (i.e., on $\Delta t$), but
instead on the spectrum.

Burst spectra can be described as $N_E\propto E^\alpha
\exp[-E/E_0]$ at low energy and $N_E\propto E^\beta$ at
high energy. The peak of $E^2N_E \propto \nu f_\nu$ occurs
at $E_p=(2+\alpha)E_0$, and thus $E_p$ is a measure of
spectral hardness. $F_T$ is most strongly a function of
$E_p$, although there is a residual dependence on $\alpha$
and $\beta$.

Therefore, detectors should be compared through sensitivity
curves in the $F_T$-$E_p$ plane.  Bursts populate this
plane: XRFs have small values of both $F_T$ and $E_p$,
while the burst hardness-intensity correlation means that
bursts will populate a band from the lower left to the
upper right of the $F_T$-$E_p$ plane.

\section{Calculation Methodology}

While the sensitivity can be calculated in detail for each
detector as a function of the source position in the FOV,
taking into account variations in the background, here I
describe simplified calculations\cite{2}.

{\bf Detector response}---I assume that the detectors are
``diagonal,'' the measured count energy is equal to the
actual photon energy (i.e., infinite energy resolution).
Since I integrate spectra and background over broad energy
bands, this is a reasonable assumption, although at high
energy a significant fraction of the photons are
downscattered.  In general I parameterize the energy
dependence of the effective area as a series of power laws.

{\bf Background}---In general I model the background as the
sum of the aperture flux plus a constant (cts s$^{-1}$
keV$^{-1}$); in some cases I use the background rates for a
given detector quoted in the literature.  The aperture flux
is the product of the cosmic background flux and $\Omega$,
the average solid angle of the sky visible on the detector
plane. $\Omega$ is calculated from formulae for rectangular
and circular detectors.

{\bf Energy bands}---I use the sets of $\Delta E$
applicable for each detector.  For proposed detectors I use
a set that optimizes the sensitivity.

{\bf Additional factors}---I also consider the fraction of
a coded mask that is open, the fraction of the detector
plane that is active, and the angle to the second most
sensitive detector when two detectors must trigger.

{\bf Factors not considered}---I do not consider the high
energy transparency of the detectors' side shields or coded
masks (except when the mask transparency is included in the
effective area).  I also do not include scattering off the
Earth or the spacecraft.  The background calculation is
crude, particularly at high energy.

\section{Applications}

\begin{figure}
  \includegraphics[height=.25\textheight]{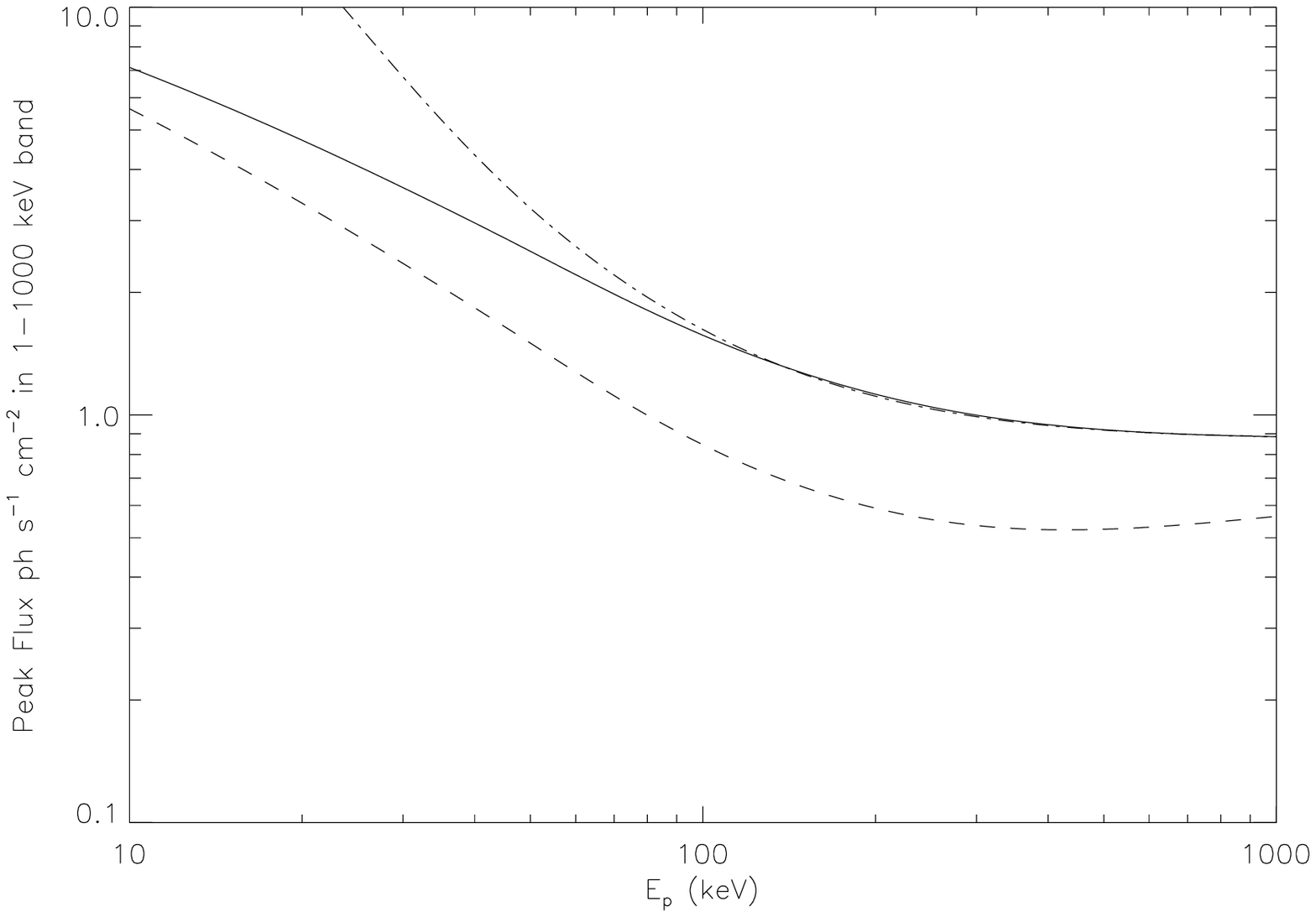}
  \includegraphics[height=.25\textheight]{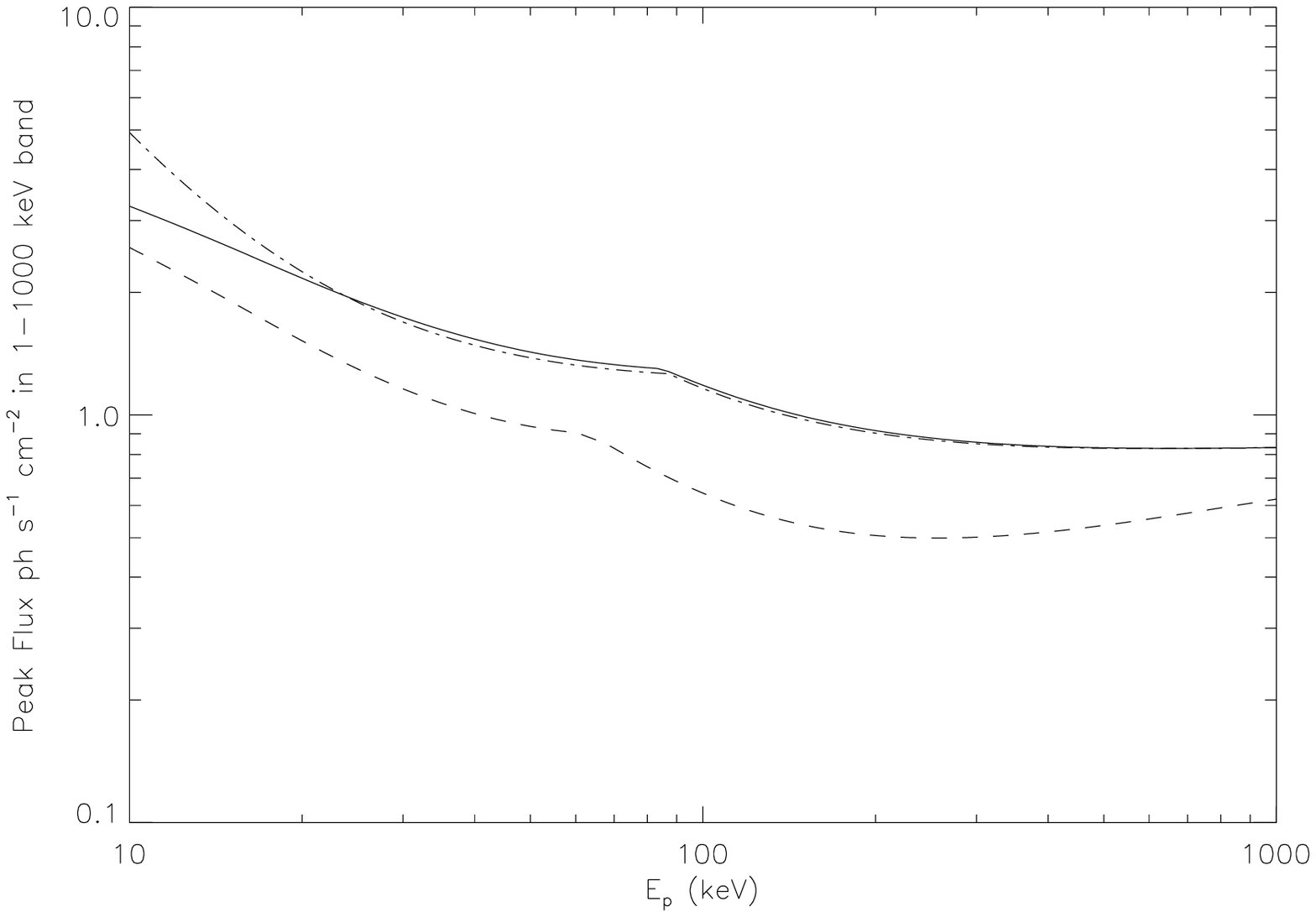}
  \caption{Sensitivity for BATSE (left) and Swift (right).
Three curves are shown: solid line---$\alpha=-1$,
$\beta=-2$; dashed line---$\alpha=-0.5$, $\beta=-2$;
dot-dashed line---$\alpha=-1$, $\beta=-3$. In all cases I
show the maximum sensitivity over the FOV, and assume
$\Delta t$=1~s.}
\end{figure}

This methodology was originally developed to compare
different detectors.  Fig.~1 compares the sensitivity of
BATSE, a set of NaI detectors, to Swift, a CZT detector.
CZT is sensitive in the 10--150~keV band as opposed to
NaI's 30--1000~keV sensitivity.  Therefore, Swift's
increase in sensitivity over BATSE will be greatest at low
energy.

\begin{figure}
   \includegraphics[height=.3\textheight]{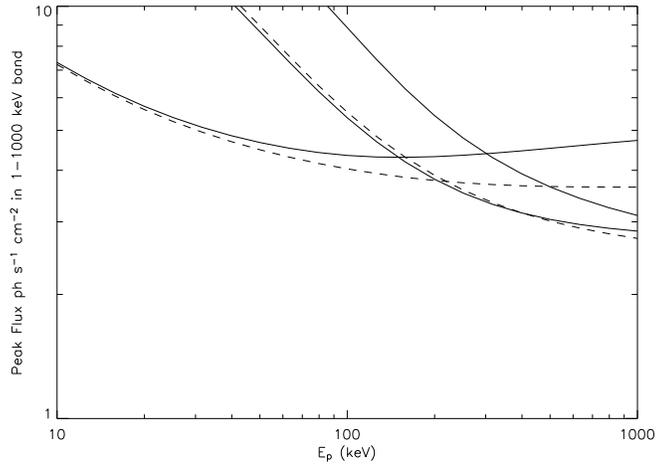}
   \caption{Sensitivity for the GBM NaI array on GLAST for
$\Delta t=1$~s and two sets of $\Delta E$; $\alpha=-1$ and
$\beta = -2$ are assumed. For the first set (solid curves,
left to right) $\Delta E$=5--100, 50--300, and
100--1000~keV while for the second set (dashed curves, left
to right) $\Delta E$=5--1000 and 50---1000~keV.}
\end{figure}

The scalloping in the Swift sensitivity curve results from
triggering on more than one $\Delta E$.  The optimal set of
$\Delta E$ can be found by comparing the sensitivities of
different $\Delta E$, as is shown by Fig.~2 from a trade
study for GLAST's GBM NaI detectors\cite{3}.  In this case
the second set of $\Delta E$ is as effective as the first,
but does not include the $\Delta E$=50--300~keV that was
BATSE's primary trigger band.

Fig.~3 shows the position in the $F_T$-$E_p$ plane that
bursts at $z=1$ with $F_T$=7.5 ph cm$^{-2}$ s$^{-1}$ and
$E_p$=30, 100, 300 and 1000~keV would have if they
originated at higher $z$.  The calculation accounts for the
narrowing of pulses at higher energy. The `+' are at
half-integral $z$ values up to $z=10$.  Also shown are the
sensitivities for BATSE, Swift and EXIST for bursts with
$\alpha=-1$ and $\beta=-2$.

\begin{figure}
  \includegraphics[height=.3\textheight]{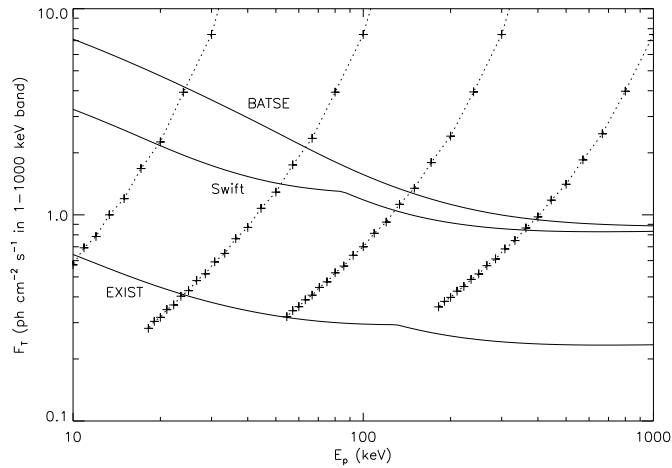}
  \caption{Tracks in the $F_T$-$E_p$ plane for identical
bursts at different $z$.  Also shown are the sensitivities
for BATSE, Swift and EXIST for bursts with $\alpha=-1$ and
$\beta=-2$.  The bursts would have
$F_T$=7.5~ph~cm$^{-2}$~s$^{-1}$ at $z=1$.}
\end{figure}

\end{document}